# Protonic Conduction Induced Selective Room Temperature Hydrogen Response in ZnO/NiO Heterojunction Surfaces

Kusuma Urs MB and Vinayak B Kamble*

School of Physics, Indian Institute of Science Education and Research Thiruvananthapuram, India 695551.

## Abstract

In this paper we show that the ionic conduction through surface chemisorbed ambient moisture leads to the remarkably high room temperature selective response towards hydrogen gas. The surface adsorbed moisture acts as surface states and shows ionic conduction, as a result of smaller size of ZnO nanoparticles of 20±5 nm. This response is enhanced remarkably i.e. from 10% to 190% for 1200 ppm $H_2$ gas when p-type NiO quasi-nanowires (width ~50 nm) are mixed with these n-type ZnO nanoparticles to form a homogenous NiO/ZnO nano-bulk *p-n* heterostructure. The maximum response is obtained for about 50-50 % composition of NiO/ZnO although it is of still n-type character. The dominant carrier type reversal from n to p type takes place at rather high NiO content of about 60-80% in ZnO, depicting dominating contribution of ZnO into the response. The parallel surface ionic current through chemisorbed moisture (surface states) has been identified as a primary factor for high sensitivity at room temperature. Thus, the presence of heterojunction barrier at the NiO-ZnO interface assisted with the surface ionic conductivity due to adsorbed moisture results in large, selective response to hydrogen at room temperature.

*corresponding author: email kbvinayak@iisertvm.ac.in phone +91-471-277 8056.

# 1. INTRODUCTION

Various hazardous gases like $H_2$, $NH_3$, CO, NO and VOC's are released every day in a very large amount from agriculture, industries and automobiles[1, 2]. Some of these gases are explosives on exposed to air after exceeding the threshold value that is as low as few parts per million, while the others are harmful air pollutants [3, 4]. Therefore, detection of these gases with high sensitivity, selectivity and response time is most needed. In this regard, various types of gas sensors such as resistive, acoustic wave, optical, electrochemical and thermoelectric are developed [4-8]. Among these, the resistive sensors using semiconducting metal oxides (SMO) are widely used because of their simplicity, cost effectiveness and ease of fabrication [9, 10]. It is evident that the transduction mechanism in resistive sensors depends on the change in the resistance of the gas sensing layer on exposure to various gases. Hence, the size, shape and composition of the sensing layer are critical in dictating the sensitivity and selectivity of these sensor devices[11]. The nano-dimensional SMOs have high surface to volume ratio which enables the active sites for gas adsorption and thus results in manifold enhancement in the performance [10, 12].

Metal oxides are first recognized as an effective gas sensing material by Bielanski and Haber[13] and Seiyama et. al.[14] about half a century ago. The first SMO sensor device manufactured by Taguchi came to the market in 1971[15]. Today, SMO sensors based on n-type materials like $SnO_2$, ZnO, $TiO_2$, $WO_3$ etc are well established for various gases. But, p-type materials are least exposed due to their relatively poor performance[16, 17]. In spite of this success as a primary sensor device choice, the SMOs are limited by their main high operating temperature (>150 °C)[9, 12]. This results in high power consumption in addition to the cost and the limited sensor deployability eventually. Moreover, operating at high temperature for the detection of flammable gases are dangerous. Hence room temperature (RT) sensors with low power consumption, high sensitivity, stability is highly desirable. In this regard, a lot of attention has been given to RT sensors[18]. Room temperature sensors based on single SMO nanomaterials in the form of nanoparticles, nanorods, nanowires, nanodiscs etc, are also well reported[11, 18] albeit with lower sensitivity. In addition, due to inadequate thermal energy to overcome large adsorption/desorption energies, the RT devices reported thus far sport high response as well as recovery times and poor selectivity. Hence, there is a constant driving force to fabricate the RT sensors which overcomes these drawbacks.

It is a proven strategy that in order to improve the sensor performance of the oxide materials, they are often functionalized with either catalytic noble metal nanoparticles, other oxides or even two-dimensional materials like Chalcogenides or graphene etc[12, 16, 19-22]. Often the oxides are also doped with several other dopant ions to tune the carrier concentration and catalytic response. This hybrid material resulting in heterojunction controls fermi energy through changes in chemical potential [16, 22, 23]. These materials of different fermi levels alignments result in the formation of either *n-n/p-p* homojunction, or *p-n* heterojunction[24, 25]. Although these strategies work, addition of noble metals is not economically feasible, synthesizing carbon materials could be tedious process with a very low probability of getting the uniform, homogenous product[19]. Hence, the best way to achieve reliable the room temperature sensors, is to improve oxide performance through junctions of different oxides like homo-or heterojunction materials which are economic and very reproducible [12, 20, 21].

Hydrogen gas is a colorless, odorless and flammable gas with 4-75% range when exposed to air. Hydrogen sensors are extremely important, mainly because hydrogen can be used as a fuel. Therefore, fast detection of the leaks even in a small amount is most needed for safety purpose [26-28]. SMO sensors of various nanostructures based on n-type materials like ZnO with high sensitivity for hydrogen sensors have been widely studied especially at high temperature. However, high temperature operation for a potential explosive gas is uncalled for. While, a *p*-type SMO such as NiO is less explored due to its low sensitivity compared to n-type counterparts[29, 30].

In this paper, we report a remarkable hydrogen response of ZnO-NiO nano-bulk heterostructure sensor which is made by facile solution methods. The $H_2$ gas response of the NiO/ZnO heterostructure thus formed shows the best performance reported so far for these (as shown in Table.1) and has been found to be greatly dominated by the surface ionic current through the chemisorbed water molecules, giving surface protonic conduction. Although this surface ionic current exists in ZnO, the presence of heterojunction barrier between ZnO and NiO improves the same through additional contribution of junction limited current control upon hydrogen exposure.

Table 1. The comparison of hydrogen response of this study with those of reported in literature so far at room temperature.

| Sr No | Sample composition | Operating temperature | Response to $H_2$ (%) | Concentration (ppm) | Response time (sec) | Ref |
|---|---|---|---|---|---|---|
| 1. | Zinc oxide/indium oxide nanorod | RT | 20 | 500 | 600 sec (10 min) | [31] |
| 2. | NiO nanosheets | 250 ºC | 160 | 100 | 900 sec (15 min) | [32] |
| 3. | Graphene/ZnO nanocomposites | 150 °C | 3.5 | 200 | 22 | [33] |
| 4. | Cd doped ZnO nanorods | 80 °C | 67 | 500 | 43 | [34] |
| 5. | $WO_3$ nanowire | RT | 9 | 20 | - | - |
| 6. | $SnO_2$ Film | RT | 4.7 | 100 | 34 | - |
| 7. | Pd/ZnO Nanorods | RT | 91.2 | 1000 | - | [35] |
| 8. | ZnO nanotubes | RT | 64.17 | 700 | 92 | [36] |
| 9. | ZnO Nanowires | RT | 8 | 121 | 29 | [37] |
| 10. | CuO/rGO/CuO | RT | 12 | 1500 | 80 | [38] |
| 11. | Pt/$Nb_2O_5$ | RT | 165 | 10 000 | - | [39] |
| 12. | NiO/$Nb_2O_5$ Nanoparticles | RT | 1.68 | 500 | 100.42 | [40] |
| 13. | Pd/$SnO_2$/rGO | RT | 50 | 10000 | - | [41] |
| 14. | $Cr_2O_3$/$Nb_2O_5$ | RT | 5.24 | 200 | 40 | [42] |
| 15. | NiO/ZnO hybrid | RT | 192 ± 20 | 1200 | 72 | This work |

## 2. EXPERIMENTAL

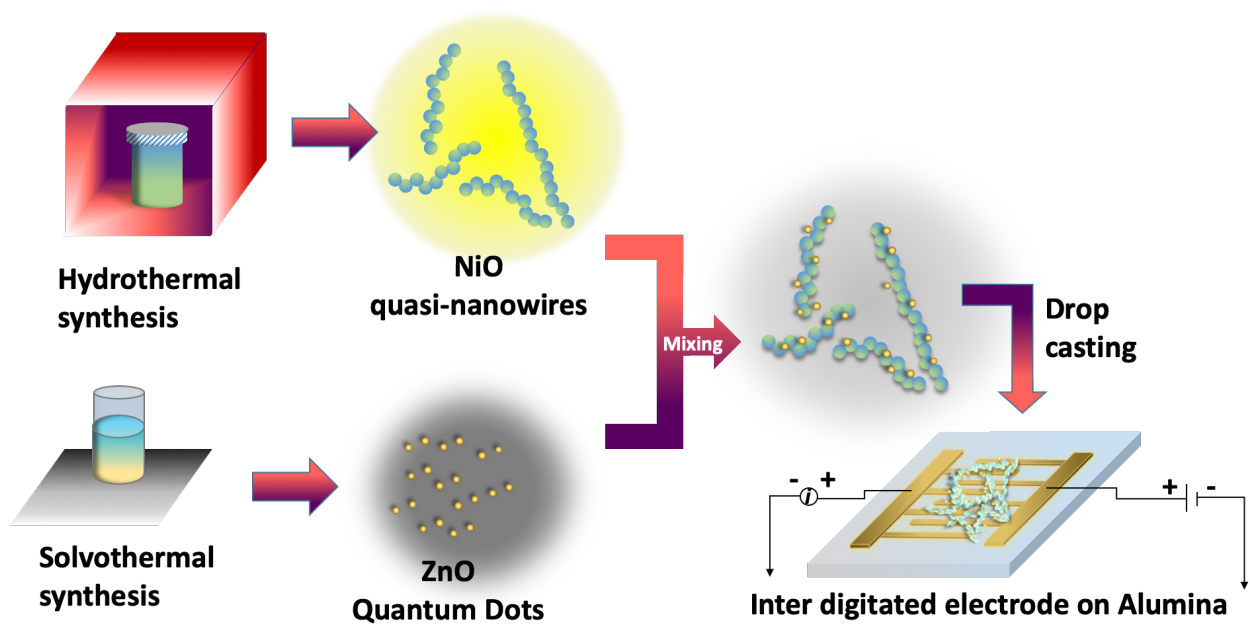

Scheme 1. The schematic diagram of the process flow used for fabrication of NiO/ZnO nano-bulk heterostructures.

### 2.1. Starting materials:

Nickel chloride hexahydrate ($NiCl_2 \cdot 6H_2O$), sodium oxalate ($C_2Na_2O_4$), zinc acetate dehydrate ($Zn(CH_3COO)_2 \, 2H_2O$), sodium hydroxide (NaOH), poly-ethylene glycol (PEG) and ethylene glycol (EG) used in this experiments are purchased from Sigma Aldrich, USA and used without any further processing.

### 2.2. Synthesis of Nickel Oxide (NiO) nanowires:

Nickel Oxide quasi-nanowires were prepared by a direct hydrothermal. A mixture of 3 mM. Nickel chloride hexahydrate ($NiCl_2 \, 6H_2O$) in $H_2O$, 1 mM Sodium Oxalate ($C_2Na_2O_4$) in 0.1 ml of PEG and 40 ml of EG were stirred for about half an hour at room temperature (25 °C). The solution is taken in a 50 ml of Teflon lined autoclave and subsequently heated to 180 °C for about 12 hours.

The product obtained was centrifuged and washed using water and ethanol. The final powder was further calcined at 500 °C for 1 hr at a heating rate of 1 °C/minute and allowed to cool to room temperature naturally.

### 2.3. Synthesis of Zinc oxide (ZnO) Quantum dots:

Zinc oxide Quantum dots were obtained by constantly stirring a mixture of 0.5 M zinc acetate dehydrate ($Zn(CH_3COO)_2\ 2H_2O$) in 50 ml of sodium hydroxide using water as a solvent at 60 °C for about 5-6 hours till a white precipitate obtained. The resulting mixture is centrifuged and washed using water and ethanol. The centrifuged product is dried at 30 °C in vacuum for about 6 hrs.

### 2.4. Synthesis of NiO-ZnO (NZ) nano-bulk heterostructure:

Various compositions of NZ were prepared of NiO/ZnO heterostructure was prepared by using of NiO quasi-nanowires and ZnO QDs as shown in Table1. The mixture was grinded using a mortar and pestle for about 1hour continuously using acetone as a solvent.

Table 1 List of compositions prepared and their sample codes.

| Sr No | Composition (NiO:ZnO wt%) | Sample code | Crystallite size (nm) |
|---|---|---|---|
| 1 | 0:100 | NiO | 27 |
| 2 | 20:80 | NZ2080 | - |
| 3 | 40:60 | NZ4060 | - |
| 4 | 50:50 | NZ | 25, 18 |
| 5 | 80:20 | NZ8020 | - |
| 6 | 100:0 | ZnO | 14 |

## 2.5. Characterizations

The as-prepared samples were characterized using Panalytical X'pert Pro x-ray diffractometer having a copper $K_\alpha$ source of 1.5418 Å in the range 20 to 70º and in step size of 0.017º. The powder samples were also characterized using Raman spectroscopy having a laser source of 532 nm. The morphology was further studied using FEI transmission electron microscope of 120 keV. The photoluminescence studies are conducted using Nanolog Horiba Scientific fluorimeter with a Xenon flash lamp. UV-Visible diffuse reflectance was studied using Perkin-Elmer Lamda 950. The gas sensing measurements was done in an inbuilt setup whose detailed can be found somewhere else[43]. 1 µL of 20 mM solution of the powders dispersed in ethanol was drop casted on to Gold Inter digitated electrodes (IDE) on alumina substrates for sensing measurements. The typical gap between the two fingers of gold IDEs were 100 µm. The UV and X-ray photoelectron spectroscopy was conducted using Omicron XPS-UPS system with Mg $K_\alpha$ source of 1253.6 eV for XPS and He $K_\alpha$ source of 21.2 eV for UPS. The energy accuracy reported are ± 0.3 eV. The system is equipped with an auto charge neutralizer however the binding energies are confirmed with reference to carbon 1s at 284.5 eV. The UPS spectra are also obtained with a bias voltage of 15V for obtaining the secondary electron cut-off.

The gas sensing measurements were performed on in house built gas sensor characterization system. The details of the system can be found in our earlier reports[22, 43].

## 3. RESULTS

X-ray diffraction of all the synthesized samples were studied to understand the crystalline phases. Fig.1(a) shows the XRD patterns of pure NiO, ZnO and 50:50 wt% NiO-ZnO (NZ) mixture. Fig.

1a(i) shows that diffraction peaks correspond to (111), (200) and (220) of NiO (JCPDS: 1313-99-1) nanowire with cubic structure having lattice parameters $a = b = c = 4.177$ Å. The XRD pattern shown in Fig 1a(ii) having indices (100), (002), (101) and other low intensity peaks ascribed to ZnO (JCPDS: 1314-13-2) corresponding to hexagonal system with the lattice *parameters $a = b = 3.249$ Å, $c = 5.206$ Å*. Moreover, ZnO peaks are considerably broadened due to the small crystallite size. While, the heterostructure composite shown in Fig. 1a(iii) confirms presence of both, NiO and ZnO peaks with no change or shift in the individual peak positions. No other peaks of impurities are observed. Thus, the coexistence of both the individual oxide nanostructures was confirmed without any significant changes in the crystallinity or phase of neither NiO nor ZnO. The crystallite size was extracted from the XRD patterns using Scherrer equation and is given in Table.1 using highest intense diffraction peaks i.e. (200) in NiO and (101) in ZnO.

The room temperature Raman spectroscopy of the samples were investigated to understand the microstructural properties and surface functional groups by studying the vibrational properties of the materials. The Raman spectra of NiO nanowire is shown in Fig. 1(b). The peaks LO (540 $cm^{-1}$) and 2LO (1090 $cm^{-1}$) corresponds to first and second order longitudinal modes of the optical phonons. The significant width of the peak LO (540 $cm^{-1}$) indicates the stretching mode of Ni-O bond and also the defects[44, 45]. Besides, several other characteristic modes of NiO are seen such as 2TO, LO+TO and the peak present at around 200 $cm^{-1}$ is due to the zone boundary phonon[44, 46].

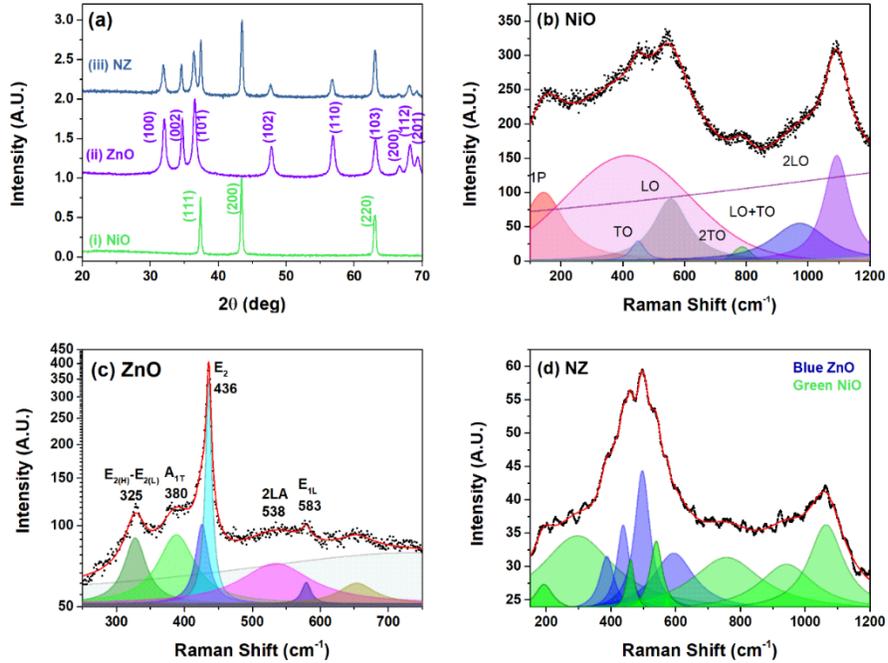

Figure 1. (a) The X-ray diffraction pattern of (i) NiO, (ii) ZnO and (iii)50-50%NiO-ZnO composite. The Raman spectra of (b) NiO, (c) ZnO and (d) 50-50% NiO-ZnO at room temperature.

These modes are significantly broadened which arises because of smaller dimensions of the crystallites. Fig 1(c) depicts the Eigen modes, $A_{1TO}$ (380 cm$^{-1}$), $E_{2H}$ (436 cm$^{-1}$), $E_{LO}$ (583 cm$^{-1}$) denotes the first order optical vibrational modes corresponds to wurtzite hexagonal structure of ZnO[47, 48]. While, $E_{2H}$-$E_{2L}$ (325 cm$^{-1}$) corresponds to second order vibration mode originated from scattered phonon at the zone boundary and 2LA (538 cm$^{-1}$) corresponds to second order longitudinal acoustic phonon vibration. Raman spectrum of NZ in Fig 1(d) shows that $E_2$ (436 cm$^{-1}$) mode of ZnO which is only Raman active is red shifted to 439±0.5 cm$^{-1}$ in NZ. It indicates the stress along c-axis of wurtzite hexagonal ZnO in a composite. The peak at 1090 cm$^{-1}$ has also been significantly broadened in the composite, which could be due to the overlapping NiO and ZnO.

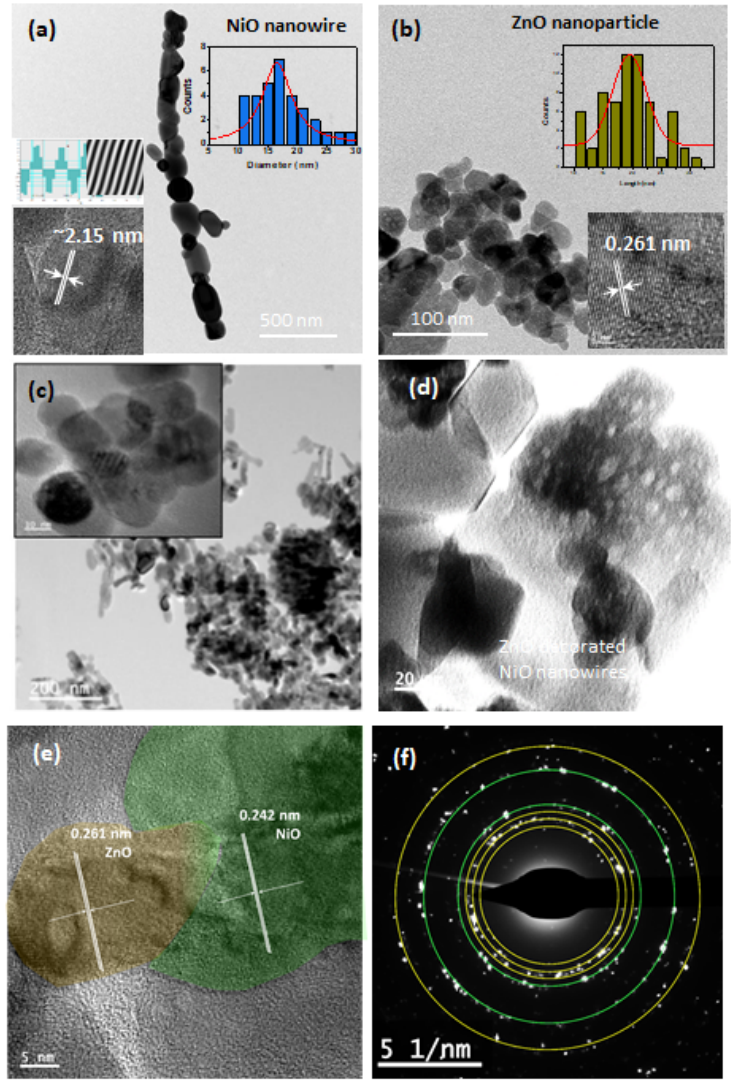

Figure 2. The Transmission electron microscope images of (a) NiO and (b) ZnO with respective HRTEM and size distribution shown insets, (c) the NZ composite (d) High resolution of NZ composite, (e) HRTEM of NZ composite showing the diffraction planes (f) SAED pattern of NZ composite.

The size and the nanostructure morphology of the samples was confirmed by Transmission Electron Microscopy (TEM). It can be seen from Fig. 2(a) that the NiO forms in a quasi-one dimensional structure. Thus, we refer it as a quasi-nanowire form. Usually the nanowires grown using vapor based techniques show single crystalline nature expect for the high aspect ratio. With the typical width (20 nm) and length (~1000 nm), here we obtain such high aspect ratio in addition to intermittent grain boundaries Further, the HRTEM in Fig. 2(a) shown inset, confirms that the NiO is a polycrystalline material with (111) and (200) diffraction planes (See S1) and that matches with the XRD pattern. The average particle size is found to be ~ 16 nm using a Gaussian fit over the size histogram. Fig 2(b) shows the TEM image of ZnO nanoparticles which are of spherical in shape and distribution of size shown in the inset of Fig. 2(b reveals that the average size of nanoparticle is 20 ±5 nm. Fig 2(c) and (d) confirms the decoration of ZnO nanoparticles onto the NiO nanowires. From HRTEM of the NZ composite presented in Fig. 2(d), (111) plane of NiO and (002) plane of ZnO can be seen that confirms the co-existence of NiO and ZnO in the NZ composite. It is further verified by the SAED pattern shown in Fig. 2(f) that displayed the different planes of NiO and ZnO.

Optical characterizations like the room temperature Photoluminescence(PL) spectra of all the samples were also investigated to understand the recombination process and to get an idea of the defect states. The spectra obtained are shown in Fig. 3(a). The presence of different types of defect states and their relative density are strongly decided by the synthesis method, size and structure of the nanomaterials. NiO and NZ samples were excited with 310 nm and ZnO samples with 360 nm. For pure NiO, a strong and intense interband emission i.e., a radiative recombination of electrons in the conduction band with the holes in the valence band is observed at 338 nm while, a very low intense and broad violet emission at 400 nm and 426 nm[49]. In case of ZnO, near band edge emission is present at 403 nm and a very less intense emission peak at 490 nm due to the Zinc interstitials and a broad green emission is due to the non-radiative recombination of electrons in the bottom of the conduction band with the electrons in the sub-bandgap of oxygen vacancies subsequently the radiative recombination with the electrons in the VB[50]. While, the PL spectrum of NZ contains both of the individual band edge emission with the band edge of NiO being slightly shifted due to the band alignment between NiO and ZnO. Moreover, there are additional peaks observed at 390 nm and 425 nm. These two peaks are attributed to the transition from metal interstitial($M_i$) and ionized metal interstitial($M_i^+$) to valence band[50] and the blue emission peak at 461 nm is due to anti-site defect of oxygen occupied in Zn lattice that acts as acceptor defects.

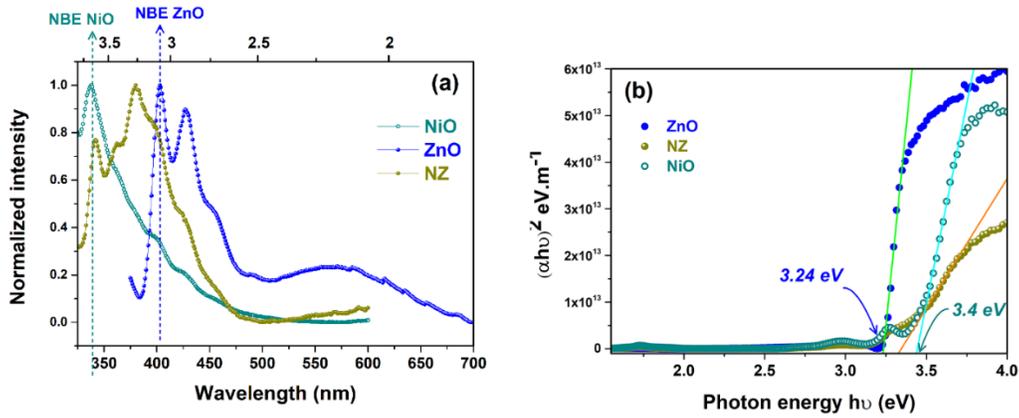

Figure 3. (a) The photoluminescence spectra and (b) the Tauc's plot of NiO, ZnO and NZ (50-50 mix) sample for estimating the optical bandgap.

.

UV-Visible absorption and diffuse reflectance spectrum of the samples were recorded to understand the band edge and optical bandgap. The optical bandgaps estimated from the tauc's plot are 3.24 eV and 3.4 eV for bare ZnO, NiO respectively and is illustrated in Fig.3(b). While, optical bandgap of the composite is 3.33 eV which clearly indicates internal electric field caused by the band bending at the junction. NiO has a broad absorption peak in the visible region that is attributed from the green color of the solution. It is reflected even in NZ sample indicate the independent contribution of NiO in the composite. The absorption edge of ZnO is at 365 nm while, that of NiO is at 320 nm. However, the absorption edge of the composite is red shifted in the wavelength to 377 nm. It is attributed to the exchange interaction between sp-d orbital band electrons and localized d electrons in Ni metal ion.

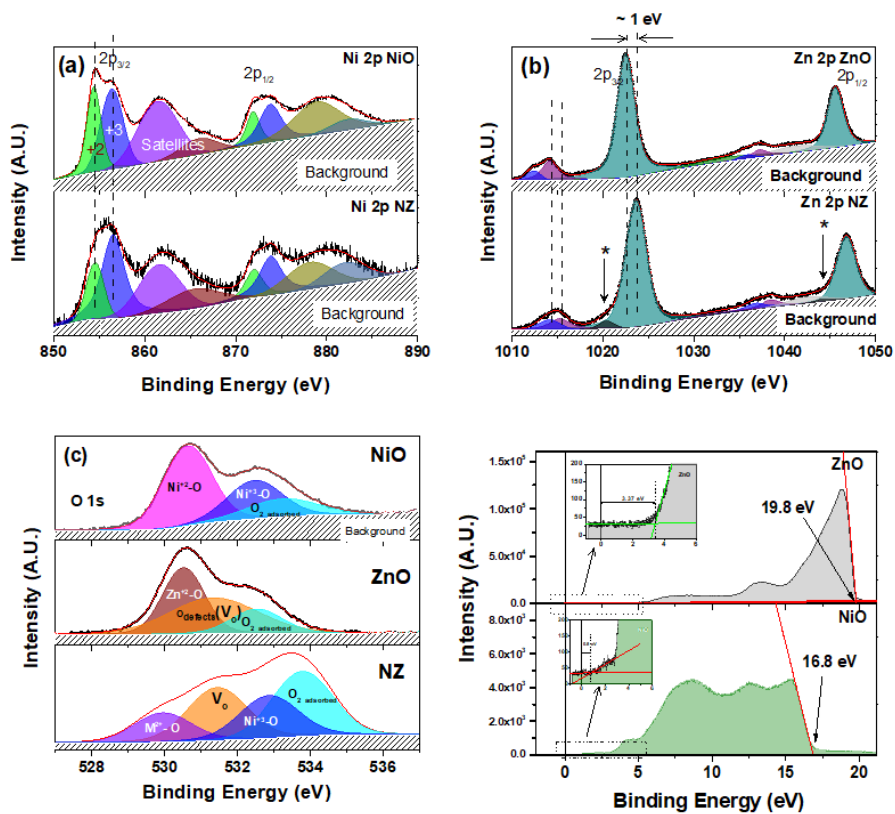

Figure 4. The X-ray photoelectron 2p core level spectra of (a) Ni 2p (b) Zn 2p and (c) O 1s from NiO nanowires, ZnO QDs, and their heterostructure composite. (d) the UV photoelectron spectra of NiO nanowires, ZnO QDs showing the valance band region.

The chemical compositional and elemental analysis of the samples were investigated using X-ray photoelectron spectroscopy (XPS) and Ultra-violet spectroscopy (UPS) to get a deeper insight about the electronic structure in composites. Fig. 4(a) shows the comparison of XPS peaks of Ni 2p in pure NiO and NZ. The spin-orbit doublet of 2p peak are resolved as $2p_{3/2}$ and $2p_{1/2}$ in accordance with spin-orbit coupling multiplicities.

Along with these two satellites for each 3/2 and 1/2 peaks are also present. When NiO is prepared in normal ambient conditions, it is bound to have metal vacancies ($V_{Ni}$) which make shallow acceptor levels[51]. These are responsible for p-type conductivity in NiO. The charge imbalance is compensated by presence of oxidized Ni ions within the NiO lattice along with vacancies.

$$3\ Ni^{2+} \rightarrow 2\ Ni^{3+} + V_{Ni} \qquad \ldots(1)$$
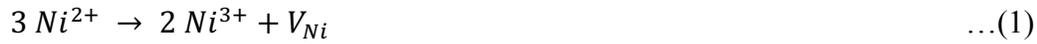

i.e.

$$NiO \rightarrow h^+ + Ni_{1-x}V_{Ni,x}O \qquad \ldots(2)$$
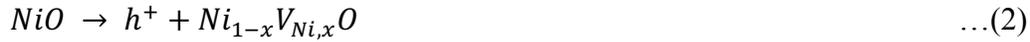

The XPS peaks of Zn in both the samples are shown in Fig 4(b). Zn 2p peaks are also resolved as $2p_{3/2}$ and $2p_{1/2}$. Zn $2p_{3/2}$ and $2p_{1/2}$ peaks in NZ (1023.6 eV) is shifted by 1eV as shown in Fig. 4(c) from that of pure ZnO (1022.5 eV). Along with 2p3/2 and 1/2, Zn intestinal peaks also present and is marked as "*" in the figure. So, the core level of ZnO has upshifted on contact with the NiO. On de-convoluting, O1s core level spectra apart from oxygen directly/covalently bonded to the metal (M-O) peaks, other peaks are also present. Oxygen stoichiometry is mainly affected by the surface groups and the interface interactions. A peak at higher binding energy is attributed for the oxygen defects. Physically adsorbed oxygen peak is present in ZnO and reflected in NZ. This confirms that the ZnO prepared using this method is very reactive to the air. The oxide peak is down shifted by 1eV in the composite due to the change in the environment at the interface. The atomic wt% is listed in the Table S1.

Ultra-violet photoelectron spectroscopy (UPS) was investigated to understand the valence band spectrum and the work-function of the samples. It can be seen from Fig.4(d) that the Fermi level ($E_F$) of NiO is at 0.8eV and ZnO is at 3.37eV which clearly depicts that the NiO is p-type and ZnO is n-type. While, the $E_F$ of NZ composite is 2.45 eV i.e in between pure NiO and ZnO. The calculated work-function values from UPS is 5.2 for pure NiO, 4.77 for pure ZnO and 4.09 for NZ. It is evident from these values that band alignment has taken place on mixing ZnO with NiO, and significantly modified the electronic structure of individual oxides in the composite.

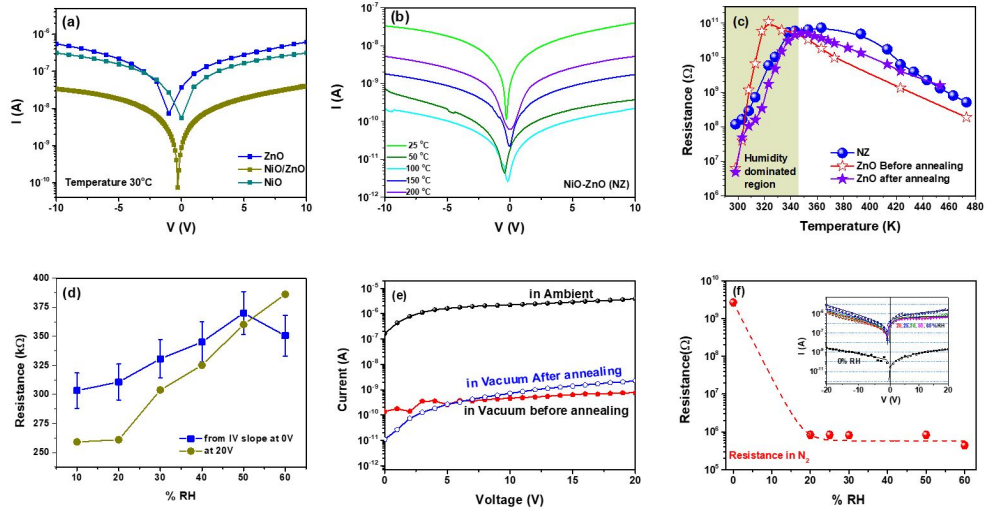

Figure 5. The Current-voltage characteristics of (a) NZ, ZnO and NiO at room temperature and (b) as a function of temperature (c) Comparison of resistance-temperature profile of ZnO before and after annealing with NZ depicting the humidity dominated region at lower temperatures (d) and (f)Variation of resistance of ZnO with different % relative humidity (RH) in ambient and in Nitrogen environment (e) I-V characteristics of ZnO in ambient, vacuum before and after annealing.

The current-voltage *(I-V)* characteristics of the NZ and pure NiO, ZnO prepared as mentioned above were measured using two-probe method at room temperature (25 °C) and at regular intervals with increasing temperature till 200 °C is shown in Fig.5 (a) and Fig.5 (b). It was observed that I-V of NiO is found to be linear and thus exhibits ohmic nature and, ZnO is schottky with gold (Fig S2). It is evident from Fig 5(a) that the resistance of the composite is more than the individuals at the same temperature and thus benefits in improving the sensor response. It can be seen from Fig. 5(c) that the NZ resistance was increasing from RT till 70 °C and then starts to decreases. It is observed from the Arrhenius plot of ZnO i.e in (Fig S3(a)) that the current at room temperature is deviated to a higher value from that of the expected one. The possible reason for the increase of conductivity could be adsorption of the environmental moisture. The water molecules adsorbed will split into $H^+$ and $OH^-$ and results in increase of ionic conductivity on the application of electric field at room temperature. The role of moisture is dominant at room temperature because the as prepared ZnO powder was not pre-heated or annealed. Thus, it can be concluded that the initial increase in resistance of ZnO and thus NZ with temperature is originated due to the decrease in the amount of conducting protons generated from the adsorbed moisture. The effect of moisture on the resistance of bare ZnO was further confirmed by varying the relative humidity of the chamber in ambient conditions and a slight increase of resistance was observed with increase in %RH. To further account the change in resistance to the humidity, ZnO resistance was measured in vacuum before and after annealing by varying the % RH. As seen from Fig.5(e), three-fold decrement in the current value in vacuum was observed to that of the ambient. The resistance value of as prepared sample decreased initially with humidity and showed almost negligible change on further inducing various % RH in nitrogen environment as seen from Fig. 5(f). Further, the activation energy ($E_a$) of the composite was extracted from the slope of ln(I) against reciprocal temperature (1000/T), the $E_a$ value for the NZ was found to be 346 meV (see Fig S3(b)).

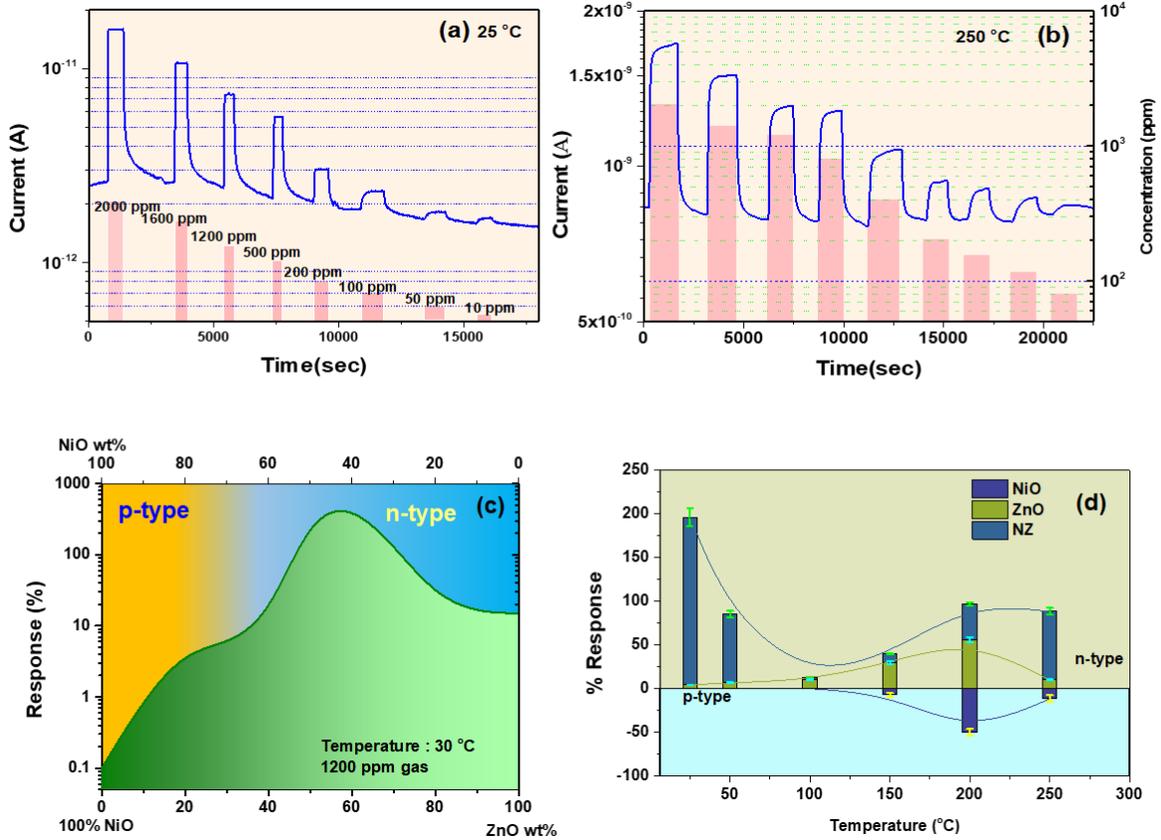

Figure 6. The response of NZ for various pulses of hydrogen gas at (a) room temperature and at (b) 250 °C. (c) The response calculated for different NiO/ZnO mixing hybrids for 1200 ppm hydrogen gas at room temperature. The switch over from p-type to n-type was observed at even 30% ZnO loading into NiO. (d) The variation of response of NiO, ZnO and NZ at different temperatures.

Usually, in conductometric sensors, the samples to be tested is coated/drop casted between two electrodes and the change in current/resistance of the sample in presence of test gas(analyte) is measured. The change in current/resistance to reach 90% of the saturation once the analyte is introduced is termed as response time and the change in current/resistance to come back to 90% of the initial value after the removal of the analyte is known as recovery time. The sensor response (*R*) for n-type and p-type is given below

$$R = \frac{I_g - I_a}{I_a} \text{ (n-type)} \quad R = \frac{I_a - I_g}{I_g} \text{ (p-type)} \qquad \ldots\ldots\ldots(3)$$

Where, $I_a$ is the current in presence of air and $I_g$ is the current in presence of analyte.

To investigate the optimum working temperature of the sensor prepared, hydrogen gas was initially used as a testing analyte. The dynamic sensor performance of all the samples was measured from RT (25 °C) to 250 °C for a wide concentration range (2000-10 ppm). The typical transient response curves shown in the Fig S5. that resistance of the bare ZnO has decreased in the presence of the hydrogen gas and that reflects in the 50:50 wt % of NZ sample though, the resistance of NiO increased in presence of hydrogen atmosphere. It can be seen from Fig. 6(a) for 1200 ppm, NZ exhibited sensor response of 10 fold higher than that of pure ZnO while, NiO didn't show any notable change at room temperature with an improved response and recovery time of around 1.5 and 25 minutes respectively. The lower detection limit(LDL) of NZ has also been improved compared to the individuals. Further, the sensor response of the composite was plotted against temperature and found that the response was higher at RT and decreases till 100 °C and then increases. Thus, there are two regions to consider to understand the sensing behavior. At lower temperatures from RT to 100 °C, ZnO dominates the sensor performance. As the temperature increases the adsorbed oxygen which is more in ZnO that is also evident from XPS decreases resulting in the number of active sites for the hydrogen adsorption and hence the response decreases till 100 °C. When the temperature further increased, NiO dominates the sensing performance in the composite and thus the response starts increases again although the overall response is n-type behavior. Hence, to investigate the phase change behavior and the senor performance, different wt% of NiO/ZnO mixtures were prepared. The response of all the mixture for 1200 ppm of $H_2$ gas at RT is presented in the Fig 6(c). It can be concluded that the mixture with 50-50 wt% showed a high sensor performance comparatively.

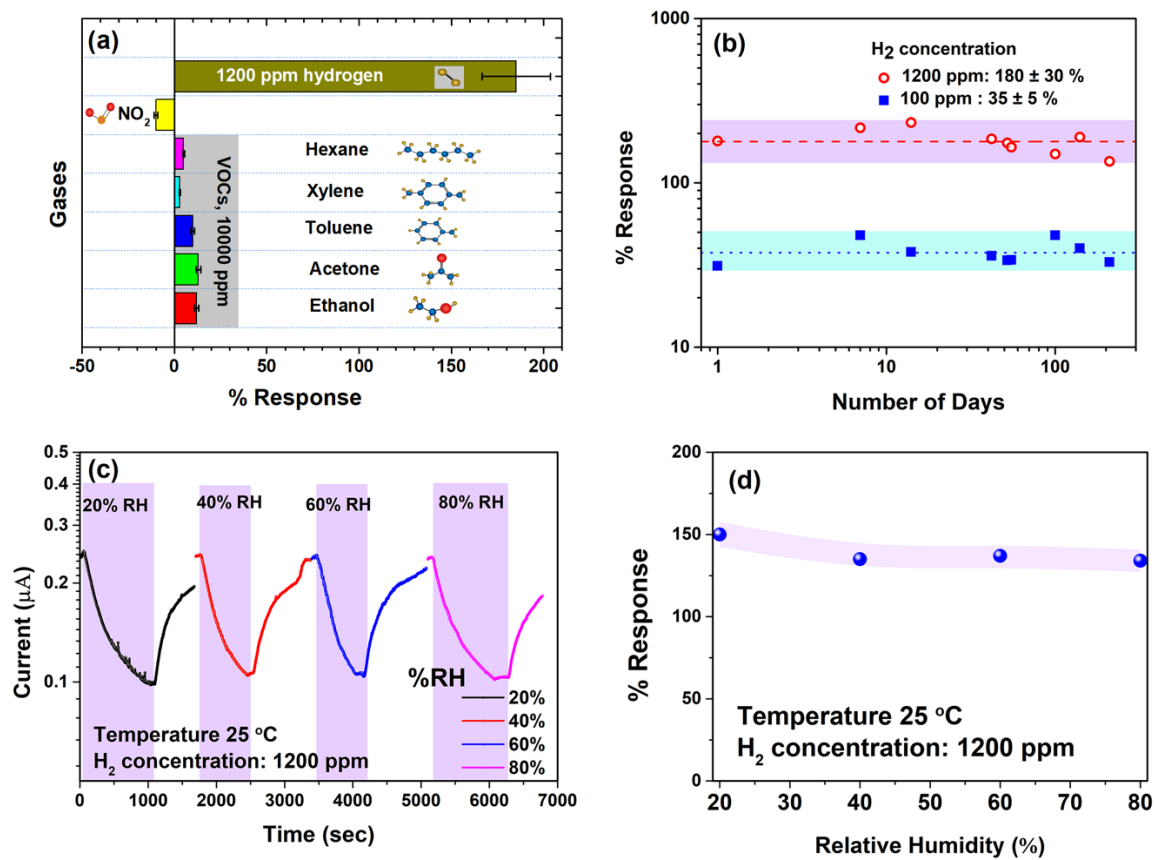

Figure 7. (a) The comparison of response of NiO/ZnO (NZ) sample for various volatile organic compounds and $NO_2$ with hydrogen at room temperature (b) The stability study of response over 200 days for high (1200 ppm) and low (100 ppm) $H_2$ concentrations. (c) The response to 1200 ppm $H_2$ at different relative humidity (%) and (d) showing humidity blind response of NiO/ZnO showing over wide humidity levels.

Selectivity of the NZ composite sample was studied at RT to see the cross-sensitivity to other gases like ($NO_2$ and VOC's) and is plotted in Fig 7(a). It is evident from the figure that the NZ prepared by this method showed a negligible response to other analytes tested compared to hydrogen gas and thus exhibits an excellent selectivity towards hydrogen gas at room temperature.

The experimental data were fitted using power law in equation (4) i.e,

$$R = AC^\beta \qquad (4)$$

Where, A and $\beta$ are the typical constant for a sensing material. For, NiO/ZnO at RT A is found to be 0.35 and $\beta$ is obtained as 0.91. Finally, the comparison of NiO/ZnO 50wt% with the individual NiO and ZnO for 1200 ppm is shown in the Fig. 6(d). The composite showed a better sensing performance as compared to the individuals. The result obtained is compared with the other in the literature and listed in Table S2 in supporting information section.

To study the stability and reproducibility of the sensor, the sensor performance is tested for almost 6 months. The response is around 190% ±20% for 1200 ppm of $H_2$ for about 5 months and is shown in Fig 7(b). The value then decreases to 130%.

Humidity effects for a wide range from 20 to 80% are studied on NZ for practical applications. Humid air is generated by passing 500 SCCM of air through a water bubbler which is heated to the temperature to get the required humidity value. It can be observed from the Fig.7(c) that the humidity has increase the conductivity of the sample and doesn't have a huge impact on the response value of a composite. Though, the value has reduced from 190 to 150 for 1200 ppm of $H_2$ and remains same for the entire range of Humidity. The slight reduction can be explained by taking into account the fact that the humid air splits into $OH^-$ and $H^+$. These ions form a bond with the oxygen vacancy ion thus result in the reduced number of oxide ion for the adsorption of hydrogen molecule.

**Sensing Mechanism**

Here, ZnO has majority carriers as electrons and NiO has holes as majority carriers. The relative positions of Femi levels in the energy bandgaps under flat band conditions are shown in Fig 8(a) and (b) before and after the matching of fermi level respectively. Thus, as shown in Fig 8(c) when the two materials are mixed, there exists a heterojunction at the interface having hole depletion layer on NiO surface and electron depletion on ZnO surface in the interfacial region.

Besides, in oxide nanoparticle, surface oxygen plays a prominent role in deciding the sensing mechanism. In case of smaller ZnO nanoparticles, there exists large surface to volume ratio and hence the number of surface uncompensated bonds is large which can be regarded as surface oxygen vacancies. Thus, the surface has $Zn^{2+}$, $O^{2-}$ and $V_o$. When ZnO is exposed to the air, oxygen molecules adsorbed on to surface vacant sites and gets ionized depending on the temperature[52]. These are typically $O_2^-$ for T < 150 °C; $O^-$ for 150 < T < 400 °C and $O^{2-}$ for T > 400 °C. The oxygen adsorption also takes place by trapping the conduction electrons from the surface which leads to the formation of additional electron depletion region and hence surface band bending. This surface band bending between two ZnO particles behaves like a barrier for the motion of electrons from one particle to the other. Similarly, NiO nanowire on exposure to air, results in the formation of hole accumulation region. Unlike n-type material's surface, the surface of p-type material, is more conducting due to hole accumulation layer.

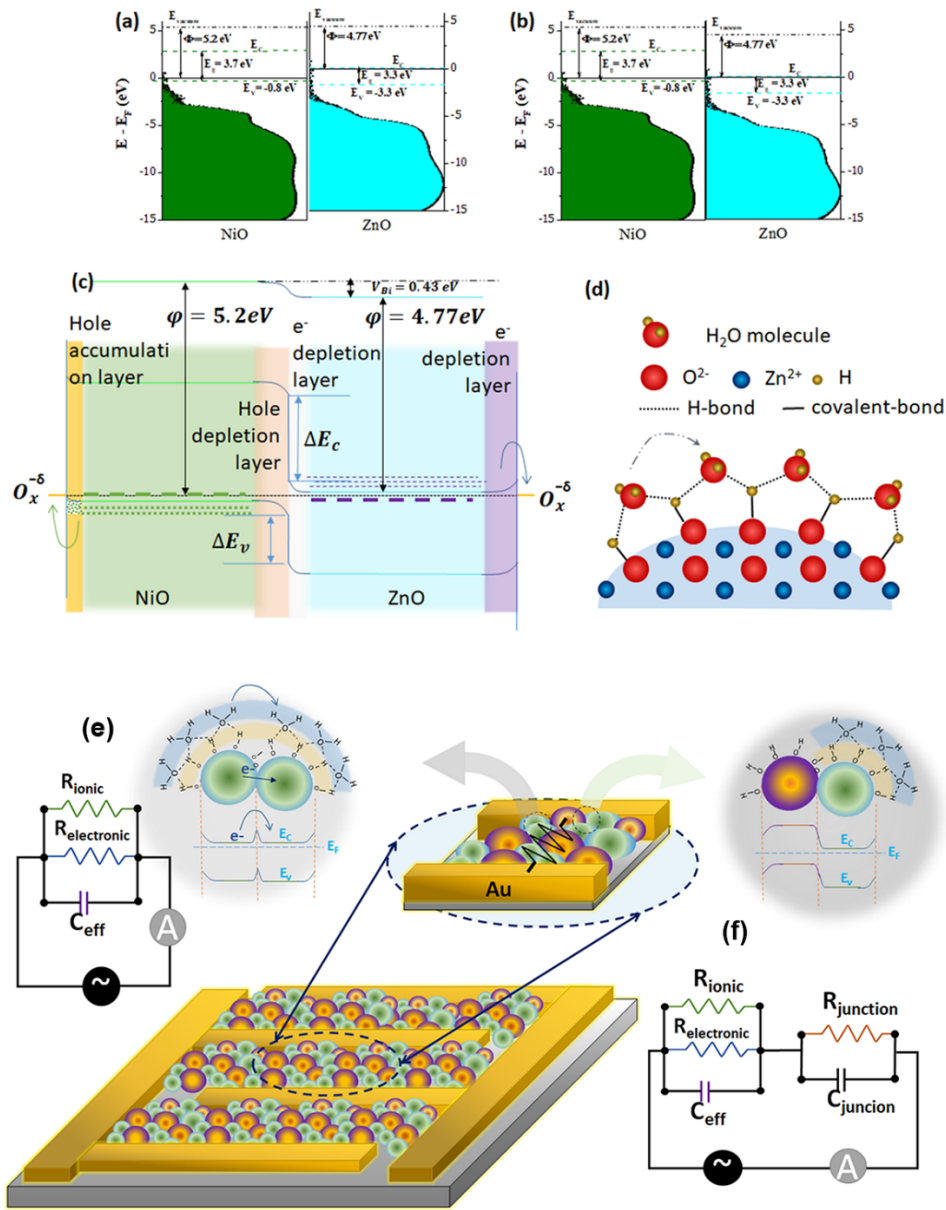

Figure 8. The density of states for NiO and ZnO from UPS studies (a) in absence and (b) upon formation of heterostructure formation. (c) The electronic band diagram of NiO/ZnO p-n junction heterostructure. (d) The schematic cartoon of moisture physisorption on ZnO surface through weak H bonds. The Schematic diagram showing the sensing mechanism and equivalent circuit of (e) only ZnO and (f) NiO/ZnO heterostructures.

When, the sensor surface is exposed to H₂ gas introduced, it gets chemisorbed in the existing surface adsorbed oxygen species. The H being a reducing gas, donates an electron to the oxide surface state forming an -OH-. This transfer of electron into surface state resulting in the release of electron trapped into it, which is subsequently given to the conduction band of the oxide. Hence, after accepting this electron, the width of the depletion region (energy barrier) decreases in n-type oxide i.e, ZnO, reducing its resistance. While, the released electron combines with excess the holes in the surface hole accumulation layer, resulting in the decrease of majority charge carriers (holes) in p-type material and hence its resistance increases. Thus, on exposure to reducing gas like H₂, CO, VOC's the conductance of n-type oxides increases while, p-type oxide decreases which can be shown as follows,

$$O_{2(gas)} + e^- \rightarrow 2O^-_{(ads)} \qquad \ldots(5)$$

$$H_2 + 2O^-_{(ads)} \rightarrow 2OH^-_{(ads)} \qquad \ldots(6)$$

$$2OH^-_{(ads)} \rightarrow 2OH_{(ads)} + 2e^- \qquad \ldots(7)$$

In addition, when there exists certain amount of moisture in the environment it also reacts with surface of ZnO to creating some more surface states. Let us write the dissociation of water molecule as follows-

$$H_2O \rightarrow H^+ + OH^- \qquad \ldots(8)$$

These H⁺ and OH⁻ ions get adsorbed on to the preadsorbed surface oxygen and vacant site respectively to form two surface adsorbed OH⁻ as shown below in eq (7),

$$Zn - O - Zn - V_o^{++} + H^+ + OH^- \rightarrow 2Zn(O-H^+) \qquad \ldots(9)$$

where, is $V_o^{++}$ is doubly ionized oxygen vacancy. Thus this can be considered as two protons chemisorbed on the ZnO surface. In the process, two positive charges inside the ZnO, which were localized on the vacancy ($V_o^{++}$) have been moved to the surface states i.e. protons. (or equivalently two electrons have been given to the lattice, increasing current due to moisture adsorption.) The next layer of moisture bonds to these H+ and O- surface species through weak hydrogen bonds as shown in Fig 8(d). There have been a number of reports who have shown existence of such surface moisture and its conduction. Milano et.al.. have particularly shown that such ionic conductivity is dominant in case of porous ZnO material and it also imparts ionic conduction to the same. Thus conduction through such granular ZnO particles in best modeled as a parallel resistance of two pathways viz, electronic conduction through lattice and ionic conduction through the surface adsorbed moisture along with their effective capacitance as shown in Fig 8(e and f).

In case of NiO nanowires added ZnO nanoparticles, the sensing mechanism involves additional contribution due to the heterojunction between two metal oxides forming p-n junction at the interface. Here ZnO is a n-type material because of presence of oxygen vacancies (donors) and NiO is a p-type material because of Ni metal interstitials (acceptors). Fig 8 (a) and (b) shows the energy band diagram of NiO and ZnO having different band gap, electron affinity and the work function. On contact, due to the formation of p-n junction at the interface of the metal oxides, it develops an internal electric field that results in the band-bending which further leads to the transfer of electrons from ZnO to NiO until the fermi level ($E_f$) equilibrates. The barrier energies for electrons and holes considering the electron affinity, band gap values obtained using UPS can be calculated as follows[53],

$$\Delta E_c = q\chi_2 - q\chi_1 = 4.77 - 1.5 = 3.27 \text{ eV} \qquad \ldots(10)$$

where, 1-NiO, 2-ZnO.

$$\Delta E_v = E_{g2} + \Delta E_c - E_{g1} = 2.87 \text{eV} \qquad \ldots(11)$$

Clearly, the conduction band offset is higher than the valence band offset. This results in the rise of resistance of heterostructure when compared to the individual NiO nanowire sensors which clearly proves the formation of p-n junction that results in the in-built electric field responsible for enhanced sensing. This effect further adds to the ionic conduction due to ZnO particles mentioned above. This is depicted in Fig 8(e and f) with additional junction impedance along with that of ZnO's native electronic and ionic conduction. Together synergistically it results in large and selective enhancement in the response at room temperature.

In case of NiO/ZnO (50 wt%) heterostructure, the conductance increases on exposure to analyte gas ($H_2$) unlike NiO nanowires. This is attributed to the increase in conductance of ZnO which is at the surface of the composite as the electron donated by the adsorbate molecule enhances the majority carriers. In a semiconductor, according to $n_0 * p_0 = n_i^2$ where $n_i$ is the intrinsic carrier concentration, $n_0$ and $p_0$ are the electron and hole concentrations. The increase in electron concentration on contact with the hydrogen enhances the electron transport across the p-n junction leading to recombination with the holes which results in the decrease of depletion width and intern, increase in the conductance. Thus, a gradual phase change from p-type to n-type behavior is observed on increasing the wt% of ZnO nanoparticles as the surface coverage increase.

4. CONCLUSIONS

In this work, the utility of hierarchical oxide hererojunctions based nanomaterials was presented with observed peculiar humidity induced effects. The sensing response enhancement for the hydrogen gas has been demonstrated at room temperature in all oxides heterojunction material prepared by an economical and simple synthesis method. Further, this composite of NiO/ZnO prepared by varying the relative proportions, showed the best response for nearly 50:50 wt% at room temperature with a dominant n-type character. The composite showed a homogenous mixing of ZnO nanoparticles with NiO quasi nanowires. The deviation of electrical conductivity of ZnO from Arrhenius behavior has been found to result from moisture induced surface ionic conductivity on granular ZnO surfaces. This has been verified by annealing in vacuum and re-exposing to humid air. The surface ionic conductivity imparts selectivity towards hydrogen through enhanced conduction as opposed to VOCs and other gases. The influence of atmospheric interfering factors has been highlighted in the study and the same has been shown to be beneficial in designing hydrogen selective, low power gas sensors.

**ACKNOWLEDGMENT**

The authors thank IISER Thiruvananthapuram Central Instrumentation Facility for access to sophisticated analytical instruments. VK would like to thank the financial support from DST Nano technology mission (DST/NM/NT/2018/124) grant from Department of Science and Technology, Government of India.

**Supporting Information**

See supporting information section for additional High resolution Transmission electron microscopy images, XPS fitting parameters, Current- Voltage characteristics and Gas sensing data.

**Author Contributions**

KU synthesized the sample and did the preliminary characterizations, device fabrication, electrical measurements and sensing studies, KU and VK analyzed the results. KU wrote the first draft. VK supervised the work. The final manuscript was written through contributions of all authors.

All authors have given approval to the final version of the manuscript.

**FUNDING SOURCES**

VK would like to thank the financial support from DST Nano technology mission (DST/NM/NT/2018/124) grant from Department of Science and Technology, Government of India.

**CONFLICT OF INTEREST**

The authors declare no conflict of interest.